\documentclass[conference]{IEEEtran}

\newif\ifincluded
\includedfalse

\usepackage{cite}
\usepackage{amsmath,amssymb,amsfonts}
\usepackage{algorithmic}
\usepackage{graphicx}
\usepackage{textcomp}
\usepackage{xcolor}
\usepackage{hyperref}
\usepackage[hyphenbreaks]{breakurl}

\begin{document}

%
\title{Invoice Financing of Supply Chains with Blockchain technology and Artificial Intelligence}

\author{\IEEEauthorblockN{Sandra Johnson}
\IEEEauthorblockA{\textit{PegaSys / ConsenSys}\\
Brisbane, Australia \\
sandra.johnson@consensys.net}
\and
\IEEEauthorblockN{Peter Robinson}
\IEEEauthorblockA{\textit{PegaSys / ConsenSys}\\
Brisbane, Australia \\
peter.robinson@consensys.net}
\and
\IEEEauthorblockN{Kishore Atreya}
\IEEEauthorblockA{\textit{Viant.io / ConsenSys}\\
Charlotte, North Carolina, USA \\
kishore.atreya@consensys.net}
\and
\IEEEauthorblockN{Claudio Lisco}
\IEEEauthorblockA{\textit{ConsenSys}\\
Sydney, Australia\\
claudio.lisco@consensys.net}
}
\maketitle
%
\begin{abstract}
Supply chains lend themselves to blockchain technology, but certain challenges remain, especially around invoice financing. For example, the further a supplier is removed from the final consumer product, the more difficult it is to get their invoices financed. Moreover, for competitive reasons, retailers and manufacturers do not want to disclose their supply chains. However, upstream suppliers need to prove that they are part of a `stable' supply chain to get their invoices financed, which presents the upstream suppliers with huge, and often unsurmountable, obstacles to get the necessary finance to fulfil the next order, or to expand their business. Using a fictitious supply chain use case, which is based on a real world use case, we demonstrate how these challenges have the potential to be solved by combining more advanced and specialised blockchain technologies with other technologies such as Artificial Intelligence. We describe how atomic crosschain functionality can be utilised across private blockchains to retrieve the information required for an invoice financier to make informed decisions under uncertainty, and consider the effect this decision has on the overall stability of the supply chain. 
\end{abstract}


\maketitle

\section{Introduction}
\label{sec:intro}
Research has identified several ways in which blockchain technology can benefit supply chains \cite{Korpela2017a, Kshetri2018b, Min2019}. Despite this active research area, certain critical challenges remain, especially around invoice financing for suppliers who do not have the `financial standing' that many tier 1 suppliers enjoy. The perceived standing (essentially the credit rating or financial status) of the supplier by the invoice financier, directly influences their risk appetite to finance the supplier's invoices, i.e. provide the supplier access, through a short-term business loan, to the amounts payable to them from unpaid invoices. The upstream supplier would be considered in a more favourable light by the financier if they were able to prove that they are part of a stable supply chain, but for competitive reasons, the retailer and manufacturer are loathe to disclose their supply chain.

We refer to the tier of a supplier within the context of the supply chain that it participates in. The number of tiers, layers or levels, that a supplier is removed from the retailer, is indicated by the tiered number. The supplier tiers are illustrated in Figure~\ref{fig:gwal} on page~\pageref{fig:gwal} . In general, the further upstream the supply chain a supplier is, the less inclined the financier is to finance their invoices. Moreover, brand awareness influences the perceived standing of the supplier, as well as the knowledge that the supplier is part of what the financier considers to be a `stable' supply chain, where stability is often based on two key factors: whether the retailer is a well known brand and whether they are financially sound. 

However, the factors that can potentially influence the stability of the supply chain are far more complex: each participant in a supply chain has a greater or lesser impact on the supply chain, and unexpected external events such as natural disasters, can greatly impact the flow of goods through the supply chain.  Despite potential adverse impacts from suppliers and external influences, the opportunity exists for an invoice financier to indirectly and positively influence the stability of a supply chain if they have the motivation to do so. On the other hand, they can also be party to the supply chain becoming unstable. 

The factors we are focussing on in this paper are two-fold: the impact on the stability of the supply chain due to the financial difficulties of one of the parties in the supply chain, and the challenge of tier 2, 3, and 4 suppliers obtaining invoicing finance. 

If the supplier experiencing financial hardship is replaced by an unreliable supplier who is unable to deliver the product on time, this will cause the supply of goods to become erratic, adversely affecting the business of other suppliers on the supply chain. Moreover, if the invoice financier is already funding invoices of another  supplier who is part of the same supply chain, then other suppliers on the supply chain, including the invoice financier, are likely to be adversely affected. For example, if an invoice financing company funds a tier 1 supplier, but does not fund a tier 3 supplier, then this not only affects the supply chain, but also the financier. Since the tier 3 supplier has difficulty in getting finance, this affects the tier 2 supplier which in turn affects the tier 1 supplier whose invoices the financier has funded.

To more clearly articulate the supply chain challenges for a supplier who is in urgent need of finance in order to fulfil the next order, or to expand their business, but who is not a tier 1 supplier, nor part of a stable supply chain, we use a fictitious supply chain use case, which has been adapted from a real world use case and simplified. 

We proffer that these challenges are not satisfactorily addressed by current solutions, but have the potential to be solved using more advanced and specialised blockchain technologies such as atomic crosschain function calls across private blockchains, and using this information to run a model to assist in the decision making of the invoice financier. We acknowledge that there may be alternative solutions to the one we present here, but we propose that our solution presents a sound and compelling reason to use atomic crosschain functionality integrated with Bayesian network modelling, an Artificial Intelligence approach, to provide a quantification of unintended consequences and the tools to make informed decisions under uncertainty, using available, but potentially partial knowledge.

\section{Example Use Case}
\label{sec:usecase}
\begin{figure}[htbp]
\begin{center}
     \includegraphics[width=\linewidth]{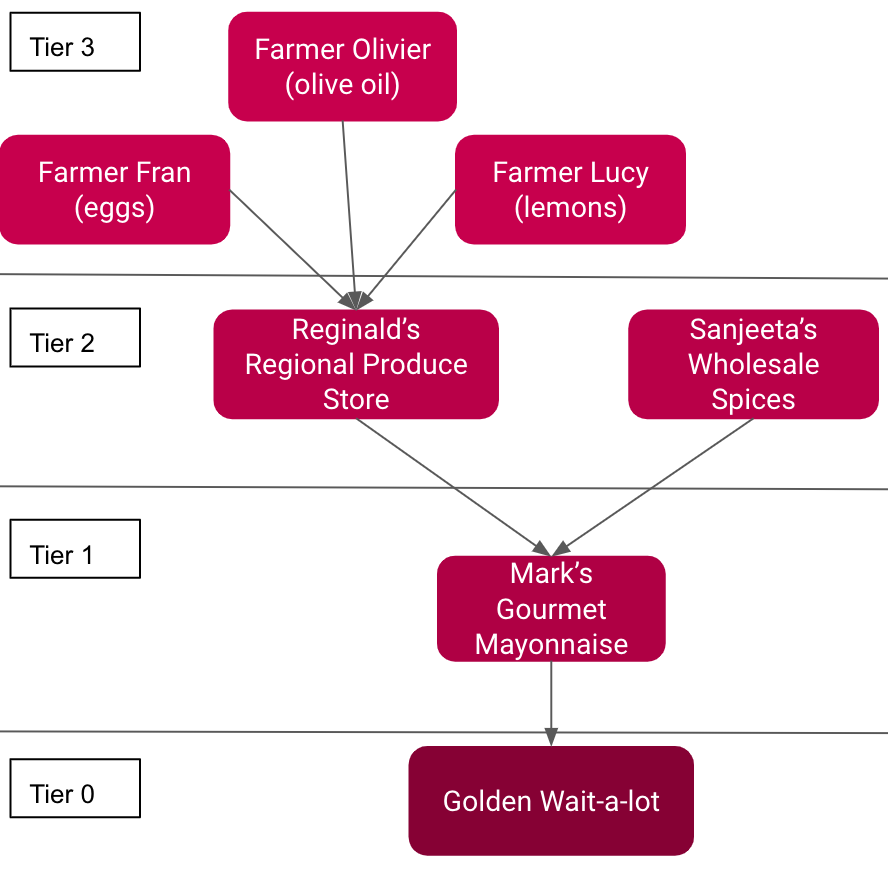}
\caption{Graphical representation of \textit{Mark's gourmet mayonnaise} range offered for sale at \textit{Golden Wait-a-Lot}}
\label{fig:gwal}
\end{center}
\end{figure}

We present a fictional scenario to demonstrate some of the potential problems that may arise in a supply chain, and to help articulate which part of the problem statement we will subsequently explore in more detail. The graphical representation of this scenario is shown in Figure~\ref{fig:gwal}.

Respected supermarket retailer, \textit{Golden-Wait-a-Lot},
has a supply chain to provide world famous \textit{Mark's Gourmet Mayonnaise} at a very affordable price to its customers. Manufacturer Mark has two supply agreements: one with \textit{Reginald's Regional Produce Store} and one with \textit{Sanjeeta's Wholesale Spices}. Mark negotiated a 60 day payment term for invoices issued by Reginald and Sanjeeta. Reginald prepares the raw produce that he receives from several farmers, according to Mark's requirements, so that they are ready to be blended, bottled and labelled by Mark for shipment to Golden-Wait-a-Lot. Reginald has negotiated a 60 day payment term with \textit{Farmer Fran} who supplies free-range eggs from her organic certified farm, a 60 day payment term with \textit{Farmer Olivier} who supplies olive oil from olive trees cultivated on his organic certified farm, and a 30 day payment term with \textit{Farmer Lucy} who supplies fresh lemons from her organically certified citrus farm. 

The success of the launch of Mark's world famous gourmet mayonnaise range at a few select stores, has exceeded Golden-Wait-a-Lot's expectations and they move quickly to expand this range into many of their other stores, negotiating increased deliveries of the product from the manufacturer. Although this appears on the surface to be a success story, it creates unintended consequences for some of the suppliers in the supply chain. 

\subsection{Problem Statement}
\label{sec:prob}
One case in point is Farmer Fran, who is unable to service the increased order that is being requested by Reginald, the regional producer, as she does not currently have the capacity to commit to this new quantity. Clearly, the other suppliers in the supply chain will have their own unique problems to solve in order to meet increased demand, but for simplicity we will explore the issues this increased order has on Farmer Fran.

Farmer Fran agreed to supply eggs which were well within the production range of her brood of hens, and offered lenient payment terms and competitive pricing with Reginald, which she expected to be more attractive than her business adversary, \textit{Farmer Eric}. Fran's reasoning was that Reginald has been known to struggle to pay invoices within 30 days, therefore 60 days would be very tempting for him, and coupled with a good price, he was more likely to accept her offer than Farmer Eric's. From her perspective it gave her a stable income so that she could slowly expand her business. Therefore she considered it a sensible strategic move. Consequently Farmer Fran is very reluctant to forfeit this great opportunity to have a guaranteed market for her organic free range eggs, and agrees to deliver the additional eggs, at the same high standard. 

The immediate challenge that faces Farmer Fran is cash flow. Personal financing (uncollateralised) is more expensive than invoice financing (collateralised). She therefore attempts to get some of her invoices financed earlier, as this will temporarily alleviate the problem by enabling her to buy additional eggs as a one-off from her friend and occasional business partner \textit{Farmer Tom}. Farmer Tom suggests that she contacts \textit{Invoice Financier Ilze} who he believes is funding Manufacturer Mark's invoices. Moreover, she does not want Farmer Eric to realise that she is in financial difficulties as he may attempt to contact the regional supplier, Reginald, to negotiate a deal to provide the additional eggs on an ongoing basis, and potentially exclude her completely from the supply chain. Therefore, it is important that Farmer Fran finds a way to prove that she is part of the supply chain to world famous Mark's Gourmet Mayonnaise (MGM), knowing MGM would not disclose their supply chain.  

For the long-term sustainability of her business Farmer Fran requires a business loan to buy the necessary infrastructure to expand her current business substantially so that she is in a position to service the increased orders on an ongoing basis. While embarking on building free-range hen houses she also needs to adhere to the regulatory requirements to ensure that she retains her organic, free-range eggs certification, which has been a long and arduous process to attain. Farmer Tom recommended that Farmer Fran applies for a business loan with \textit{Lender Luke}, who is renowned for being discreet and trustworthy in business dealings. Farmer Fran is also following up on several leads from friends for invoice financiers, including Invoice Financier Ilze, to finance her outstanding invoices, but she is not having much success in pursuing invoice financing, as these companies are notoriously averse to financing invoices from tier 2, 3 and 4 suppliers.

This scenario presents several challenges for Farmer Fran. We group the various events and requirements under two overarching problem statements, P1 and P2:

\begin{enumerate}[label=P.1]

\item \textit{Access to funds to fulfil the increased order within the requested timeframe} \\
To provide the additional eggs for the next order, Farmer Fran needs to:
\begin{enumerate}
	\item Sign an agreement with Reginald's Regional Produce for the additional delivery of eggs.
	\item Secure invoicing finance for her unpaid invoices ahead of the scheduled payment date.
	\item Source organic, free-range eggs from an alternative supplier, e.g. Farmer Tom, as a one-off transaction.
	\item Ensure data confidentiality of the details of her contract with Reginald's Regional Produce Store, so that:
	\begin{itemize}
		\item Farmer Eric does not know who she is supplying with eggs, nor the price and payment terms she negotiated.
		\item Farmer Tom does not know the payment terms and price she negotiated.
	\end{itemize}
	\item Ensure data confidentiality of her negotiations and contact with invoicing finance companies, so that:
	\begin{itemize}
		\item Farmer Eric is not aware of her cashflow problem.
		\item Other suppliers in the supply chain are not aware of Farmer Fran's cashflow problem.
	\end{itemize}	
\end{enumerate}
\vspace{5pt}
\item \textit{Loan to expand organic free-range eggs business to service increased demand on an ongoing basis}  \\
Farmer Fran ideally needs a loan to purchase the infrastructure and stock to expand her organic free-range eggs business, as well as an on-going loan agreement with an invoice financing company to fund her unpaid invoices to release funds for her to use as required.
For example, Farmer Fran knows that the egg production of her brood varies from day to day, and she will therefore occasionally need to supplement her supply of eggs with additional eggs from another source in order to fulfil her order. Moreover, she needs to ensure that all infrastructure projects adhere to the organic and free range specifications to maintain her certification.

\begin{enumerate}
	\item Secure loan for construction of hen houses and purchase of pullets
	\begin{itemize}
	\item Loan company will perform various checks, e.g. Farmer Fran's capacity to service the loan, to repay it in the requested time-frame, credit rating.
	\item Provision of documentation such as bank statements and business accounts.
	\end{itemize}
	\item Ensure confidentiality of her loan application, amount, terms and conditions.
\end{enumerate}

\end{enumerate}

In this paper we focus primarily on invoice financing of suppliers further upstream in the supply chain, i.e. tiers 2, 3 and 4. For this reason, and to limit the scope of the case study we describe an approach which addresses the first problem, P1. This is also the immediate challenge facing Farmer Fran, i.e. fulfilling the next order. Problem P2 is no less critical, but arguably not as imminent as P1. Despite focussing on P1, the approach we describe in Section~\ref{sec:methods}: Methods, can readily be applied to problem P2. 

There are a couple of interesting and concerning implications of invoice funding decisions by a finance company. We described these in the Introduction, Section~\ref{sec:intro} on page \pageref{sec:intro}, viz. the effect it may have on the finance company itself, and the overall stability of the supply chain. 

\section{Methods}
\label{sec:methods}
We describe an approach that could be implemented in smart contracts on private blockchains as part of a decision making tool for invoice financiers regarding the funding of invoices of suppliers who are further upstream in the supply chain, i.e. closer to the sourcing of the raw products being processed and sold to customers. Using a combination of Bayesian Networks (BNs) \cite{Johnson2012a, Holt2018}, which is an approach to Artificial intelligence (AI), and crosschain function calls across private blockchains \cite{Robinson2019b} we can enable financiers to make informed funding decisions under uncertainty. 

The private blockchains are therefore required to have smart contract capability and atomic crosschain functionality \cite{Robinson2019b}. For example, Pantheon, an Ethereum Java Client designed for enterprise needs \cite{PegaSys}, has smart contract capability and privacy groups which assist in providing data confidentiality \cite{PegaSys2019}. A privacy group consists of the entities involved in a transaction, who are the only entities able to view the details of the transaction. Moreover, the participant list for that transaction will be hidden from all other parties. However, crosschain functionality is not yet available in Pantheon, and a proof of concept is currently being developed by the Sidechains Research Team at PegaSys/ConsenSys.

In Section~\ref{sec:bn} we develop a Bayesian network model which will use the information from the crosschain function calls described in Section~\ref{sec:sc} to calculate the probability of whether the perceived risk of providing finance to the supplier is acceptable or unacceptable, the most probable decision of the invoice financier based on the perceived risk and the probability of the stability of the supply chain of the supplier who is applying for finance, depending on supplementary information being available, such as whether the invoice financier already funds invoices of another supplier in the supply chain.

\subsection{Bayesian network modelling}
\label{sec:bn}
The AI approach we use is an object oriented Bayesian network (OOBN) modelling technique.  Bayesian networks (BNs) are probabilistic graphical models \cite{Kyburg1991,Pearl2014}, and the process of building the model is typically an iterative process \cite{Marcot2007, Johnson2010}. We present an initial model where we consider some of the key pieces of information that contribute to making an informed decision about invoice funding. BNs are increasingly integrated with other modelling techniques and systems \cite{Johnson2012a, Marcot2019}, and we propose an example of combining blockchain technology with BN models for invoice funding of suppliers in a supply chain.

A BN is constructed as a directed acyclic graph (DAG) with nodes (ellipses) representing the key factors and edges (arrows) showing the dependencies between nodes \cite{Marcot2007, Pearl2014}. Conditional probability tables are associated with each of the nodes in the network \cite{Marcot2007, Pearl2014}. These probabilities are typically constructed from expert knowledge if insufficient data are available, or may be learnt from data if sufficient data are available \cite{Heckerman1995, Korb2010}. Moreover, the structure of the BN can be also be learnt from the data \cite{Heckerman2008}, but care needs to be taken as association between different factors do not imply causality \cite{Korb2010} and often sufficient expert knowledge is available to determine an overall initial causal structure of the model, which can then be evaluated and updated as new information becomes available \cite{Kjaerulff2008, Korb2010}. Once the model is constructed, information or knowledge, referred to as `evidence', may be entered in the model and this evidence is propagated through the model, yielding updated probabilities based on the evidence entered \cite{Kjaerulff2008}.  

\begin{figure}[ht]
\begin{center} 
  	 \includegraphics[width=0.6\linewidth]{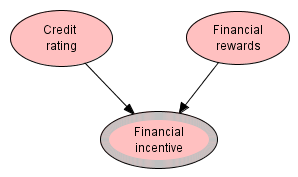}\\
	 (a) Financial incentive OOBN sub-model \\
	 \includegraphics[width=0.9\linewidth]{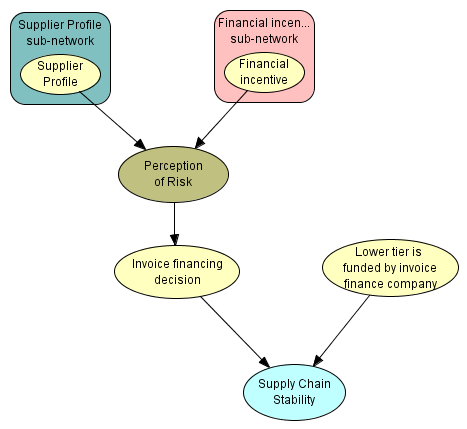}\\
	 (b) Invoice financing and Supply Chain stability OOBN \\
	  \includegraphics[width=0.5\linewidth]{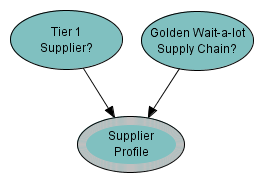} \\
	  (c) Supplier profile OOBN sub-model
  	\caption{Bayesian network model of invoice financing for a supplier and the stability of the supply chain}
	  \label{fig:oobn}
  \end{center}
\end{figure}

Figure~\ref{fig:oobn} (a) and (c) shows two OOBN sub-networks constructed using BN software, Hugin \cite{HuginExpert2019}: \textit{Financial incentive sub-network} and \textit{Supplier Profile sub-network}, respectively, which are included in the overall model Figure~\ref{fig:oobn}~(b) with the output nodes of interest: \textit{Invoice financing decision} and \textit{Supply Chain Stability}. Table~\ref{tab:nodes} contains a list of the nodes in the BN models and the states of the nodes. For example, node \textit{Financial incentive} can be in two states: \textit{Compelling} or \textit{Not compelling}. When the BN model containing this node is run, it will generate probabilities for both states of the node, i.e. what is the probability that the \textit{Financial incentive} node will be in the \textit{Compelling} state or \textit{Not compelling} state. The various scenarios and results of running the OOBN models are discussed in the Results section on page \pageref{sec:results}. 

\begin{center}
\begin{table}[ht]
\caption{Bayesian network models for the example use case}
\label{tab:nodes}
  \begin{tabular}{lll}
  \hline 
  \textit{BN model} & \textit{Node} & \textit{Node states}\\
 \hline
Supplier &  Tier 1 Supplier? & Yes, No\\
\cline{2-3} 
Profile & Golden Wait-a-Lot & Yes, No\\
 & Supply Chain? & \\
 \cline{2-3} 
 & Supplier Profile & Low Risk, High Risk \\
 \hline
 Financial  & Credit rating & Passed, Failed \\
\cline{2-3} 
 incentive & Financial rewards & Additional, Standard \\
 \cline{2-3} 
 & Financial incentive & Compelling, \\
 & & Not compelling \\
 \hline
  Overall & Supplier Profile & Low Risk, High Risk \\
 \cline{2-3}
 model & Financial incentive & Compelling, \\
 & & Not compelling \\
\cline{2-3}
 & Perception of risk & Acceptable risk, \\
 & & Unacceptable risk\\
 \cline{2-3}
 & Invoice financing & Fund supplier \\
 & decision &  invoices, Do not \\
 & & fund supplier \\
  & & invoices \\
 \cline{2-3}
 & Lower tier is funded & Yes, No \\
 & by invoice finance & \\
 & company &\\
 \cline{2-3}
 & Supply Chain & Stable, Unstable \\
 & Stability &  \\ 
  \hline
  \end{tabular}
\end{table}
\end{center}

\subsection{Private Blockchains}
\label{sec:sc}
The blockchain technology design for the example scenario described in Section~\ref{sec:usecase} on page~\pageref{sec:usecase} consists of several private blockchains with participants from the supply chain, national credit reporting bodies, and organic and free-range certification authorities. A possible configuration of blockchains and participants are outlined in Table \ref{tab:pb} on page \pageref{tab:pb}. For clarity we will concentrate mainly on Farmer Fran and her application for invoicing finance, which are sufficient to demonstrate cross function calls and the interaction with the BN decision making tool running on the Financier's computer system, and other activities that do not directly relate to the funding request, are not included in the exposition of private blockchains and atomic function calls, but are mentioned to raise awareness that there are many other considerations that need to be taken into account.\\

\begin{center}
\begin{table}[ht]
\caption{Private Blockchains for Invoice Financing Problem Scenario P1}
\label{tab:pb}
  \begin{tabular}{ll}
  \hline
 \textit{Private Blockchain} & \textit{Participants}\\
 \textit{identifier} & \\
 \hline
 T0T1  &  Golden Wait-a-Lot (tier~0), \\
 & Manufacturer Mark (tier~1)\\
\hline
T1T2 & Manufacturer Mark (tier~1), \\
 & Reginald's Regional Produce Store (tier~2), \\
 & Sanjeeta's Wholesale Spices (tier~2) \\
 \hline
 T2T3 & Reginald's Regional Produce Store (tier~2), \\
  & Farmer Fran (tier~3), \\
  & Farmer Olivier (tier 3), \\
  & Farmer Lucy (tier~3) \\
\hline
 T3Fin & Farmer Fran (tier~3), \\
   & Invoice Financier Ilze \\ 
\hline 
T3T3 &  Farmer Fran (tier~3), \\
& Farmer Tom (tier~3) \\  
\hline
Fin & Invoice Financier Ilze \\
 \hline
 Cert & Certification authority \\
  & (organic \& free-range) \\
    \hline
  \end{tabular}
\end{table}
\end{center}

The information that is required from the private blockchains, some of which have been specifically set up as a result of Farmer Fran requesting invoice financing to get access to cash so that she can commit to providing the additional organic free-range eggs to Regional Producer, Reginald, by the specified date, involves several transactions (reading and writing), some of which need to be atomic crosschain function calls. A detailed description of the blockchains and information flow through the system is outlined in Section~\ref{sec:xcresults} on page~\pageref{sec:xcresults}.

\section{Results}
\label{sec:results}

\subsection{BN models}
\label{sec:bnresults}
Each of the OOBN sub-models can be run independently, and as part of the overall BN model (Figure~\ref{fig:oobn} (b)). Various scenarios are described in this section and the output from running the scenarios is shown as probabilities that a node is in a particular state, given the evidence that has been entered into the model.

\subsubsection{Supplier Profile OOBN}
\begin{figure}[ht]
\begin{center} 
  	 \includegraphics[width=0.55\linewidth]{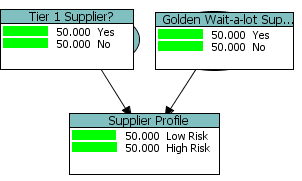}\\
	 (a) Running model with no evidence entered \\
	 \includegraphics[width=0.55\linewidth]{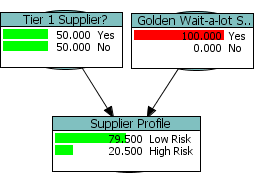}\\
	 (b) Running model with knowledge that supplier is in a Golden Wait-a-Lot supply chain\\
	  \includegraphics[width=0.55\linewidth]{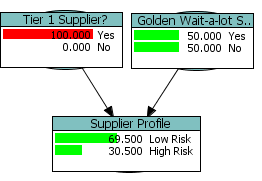} \\
	  (c) Running model with knowledge that the supplier is a tier 1 supplier \\
	  \includegraphics[width=0.55\linewidth]{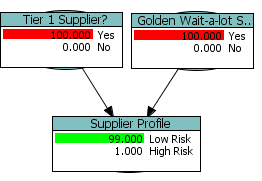} \\
	  (d) Running model with knowledge that the supplier is a tier 1 supplier in a Golden Wait-a-Lot supply chain \\
  	\caption{Supplier Profile Bayesian network model with various scenarios}
	  \label{fig:supplieroobn}  
\end{center}
\end{figure}

We ran this model using the following scenarios. The output is shown as probabilities in Figure~\ref{fig:supplieroobn} on page \pageref{fig:supplieroobn}:
\begin{enumerate}[label=(a)]
\item No evidence 
\item Supplier is part of a Golden Wait-a-Lot supply chain
\item Supplier is a tier 1 supplier
\item Supplier is part of a Golden Wait-a-Lot supply chain, and a tier 1 supplier
\end{enumerate}
 
The main output node that we are interested in, \textit{Supplier Profile}, is an input node into the overall model. With no information entered into the OOBN (scenario (a)) we observe that the supplier profile has uniform probability and is uninformative. If we enter evidence that the supplier is known to be part of a Golden Wait-a-Lot (GWaL) supply chain (scenario (b)) then the perceived profile of the supplier is that they are most likely to be low risk with a probability of 79.5\%. If we know that the supplier is a tier 1 supplier, but we have no information regarding their participation in a GWaL chain (scenario (c)), then the most likely outcome is still that the supplier is low risk, but with a probability of 69.5\% which is less than the knowledge of being in a GWaL supply chain. If we enter evidence that the supplier is both a tier 1 supplier and in the GWaL supply chain (scenario (d)) then it is highly likely, 99.0\%, that the supplier is low risk to the financier.

\subsubsection{Financial incentive OOBN}
\begin{figure}[ht]
\begin{center} 
  	 \includegraphics[width=0.55\linewidth]{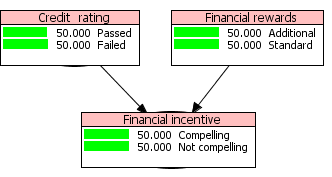}\\
	 (a) Running model with no evidence entered \\
	 \includegraphics[width=0.55\linewidth]{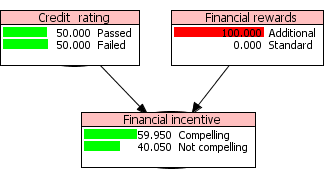}\\
	 (b) Running model with evidence that supplier is offering additional financial rewards\\
	  \includegraphics[width=0.55\linewidth]{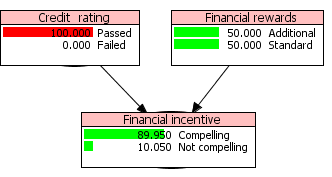} \\
	  (c) Running model with evidence that the supplier has passed the credit check \\
	  \includegraphics[width=0.55\linewidth]{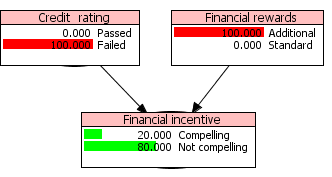} \\
	  (d) Running model with evidence that the supplier failed the credit check, but is offering additional financial rewards\\
	   \includegraphics[width=0.55\linewidth]{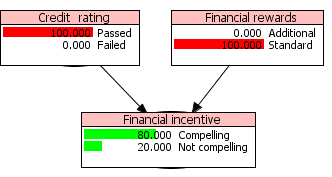} \\
	  (e) Running model with evidence that the supplier passed the credit check, and is offering just the standard financial rewards\\
	  \includegraphics[width=0.55\linewidth]{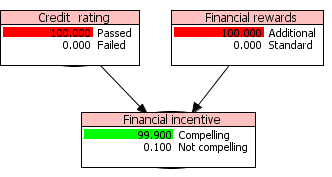} \\
	  (f) Running model with evidence that the supplier passed the credit check, and is offering additional financial rewards\\
	  \caption{Financial Incentives Bayesian network model with various scenarios}
	  \label{fig:incentivesoobn}
    \end{center}
\end{figure}

We ran this model using the following scenarios. The output is shown as probabilities in Figure~\ref{fig:incentivesoobn} on page~\pageref{fig:incentivesoobn}:
\begin{enumerate}[label=(a)]
\item No evidence 
\item Supplier is offering additional financial rewards, e.g. offering above the normal discount rate for early payment
\item Supplier passed the credit check
\item Supplier failed the credit check, but is offering additional financial rewards
\item Supplier passed the credit check, and is offering only the standard financial reward
\item Supplier passed the credit check, and is offering additional financial rewards
\end{enumerate}

The main output node that we are interested in, \textit{Financial incentive}, is an input node into the overall model. With no information entered into the OOBN model (scenario (a)) we observe that the \textit{Financial incentive} output node has uniform probability and is uninformative. If we enter evidence that the supplier is offering better than the standard financial rewards to the invoice financier, but without knowledge of their credit rating (scenario (b)) then the probability that the incentive is compelling to the financier is 60.0\%. There is a greater probability of a compelling incentive to the financier (90.0\%) to fund invoices of the supplier, if the supplier passed the credit rating assessment, despite the financier not having any knowledge of the financial rewards they are offering (scenario (c)). If we consider being able to provide evidence about both the credit rating assessment and the financial rewards, we discover that the probability of the financial incentive being compelling is only 20.0\% if the supplier failed the credit rating assessment, even if they are offering additional financial rewards to the invoice financier. On the other hand, if the supplier passed the credit rating test, and is only offering the standard financial reward expected by invoice financiers (scenario (e)), then the incentive to finance them is 80.0\% compelling. The best case scenario, (f), is a supplier who has passed the credit rating assessment and is offering additional financial rewards. In this case they present a very highly compelling financial incentive (99.0 \% probability) to the invoice financier to fund their invoices. 

\subsubsection{Overall OOBN model}
The overall model has input nodes from the two sub-networks so that evidence entered into these sub-networks will flow through to the overall BN model. 

\begin{figure}[ht]
\begin{center} 
  	 \includegraphics[width=0.7\linewidth]{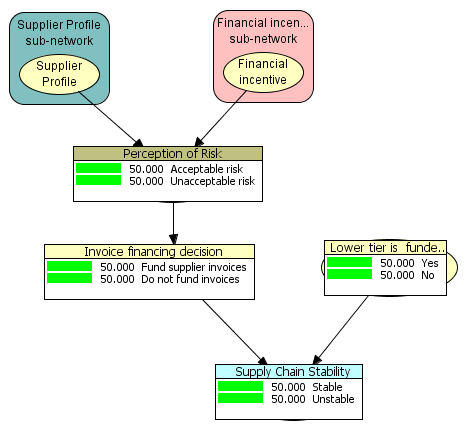}\\
	  (a) Running model without any evidence entered into the BN\\
	  \includegraphics[width=0.7\linewidth]{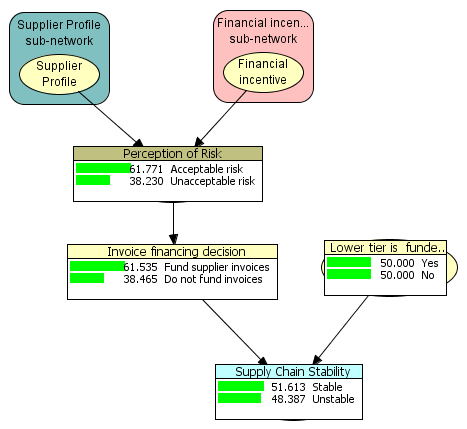}\\
	  (b) Supplier is in the GWaL supply chain\\
	   \includegraphics[width=0.7\linewidth]{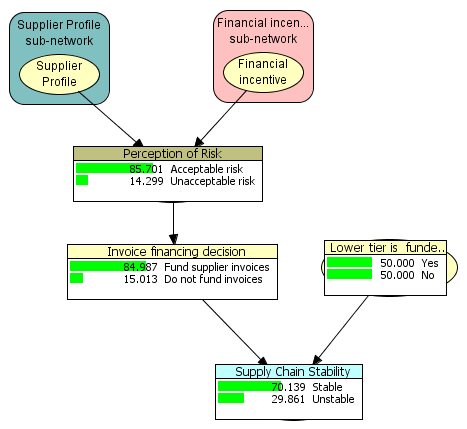}\\
	  (c) Supplier is in the GWaL supply chain and passed the credit rating assessment
  	\caption{Overall Bayesian network model with various scenarios}
	  \label{fig:bnout}
  \end{center}
\end{figure}

\begin{figure}[ht]
\begin{center} 
	  \includegraphics[width=0.7\linewidth]{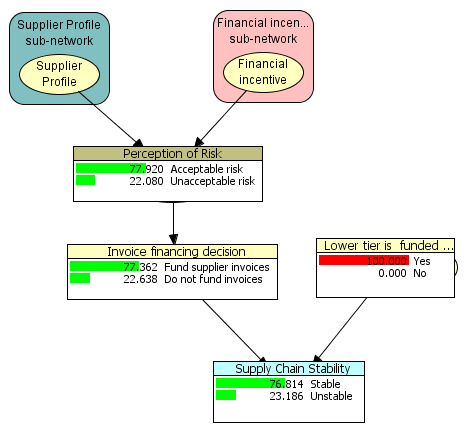}\\
	  (d) Supplier is in GWaL supply chain, passed credit check, is not a tier 1 supplier and the lower tier is funded by the invoice finance company \\
	   \includegraphics[width=0.7\linewidth]{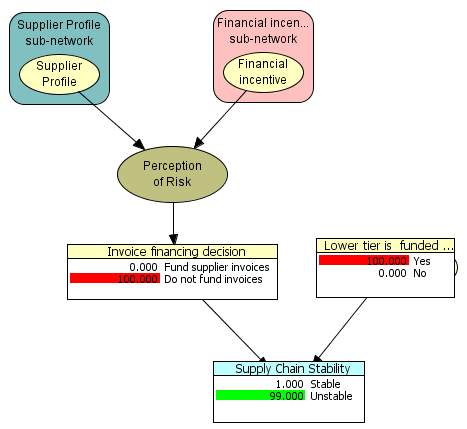}\\
	  (e) Decision was made not to fund the supplier (e.g. tier 2 or 3), although the lower tier (e.g. tier 1) is funded by the financier
  	\caption{Overall Bayesian network model with various scenarios}
	  \label{fig:bnout2}
  \end{center}
\end{figure}

We ran the overall model using the following scenarios. The output is shown as probabilities in Figure~\ref{fig:bnout}~and~\ref{fig:bnout2} on page~\pageref{fig:bnout}:
\begin{enumerate}[label=(a)]
\item No evidence 
\item Supplier is in a Golden Wait-a-Lot supply chain
\item Supplier is in a Golden Wait-a-Lot supply chain and passed the credit check
\item Supplier is in a Golden Wait-a-Lot supply chain, has passed the credit check, is not a tier 1 supplier (e.g. a tier 2 or 3) and the lower tier (e.g. tier~1) is funded by the invoice finance company
\item Decision was made not to fund the supplier (e.g. tier 2 or 3), although the lower tier (e.g. tier 1) is funded by the financier
\end{enumerate}

 With no information entered into the OOBN model (scenario (a)) we observe that the nodes of interest, \textit{Perception of Risk}, \textit{Invoice financing decision} and \textit{Supply Chain Stability} have uniform probability and is uninformative.
If it is known that the supplier is part of a Golden Wait-a-Lot supply chain (scenario (b)), then the risk profile of the supplier is considered an acceptable risk with a probability of 61.8\%, Figure~\ref{fig:bnout}~(b) on page~\pageref{fig:bnout}. If we also include evidence that the supplier passed the credit rating assessment (scenario (c)), then the perception of acceptable risk increases to 85.7\%, Figure~\ref{fig:bnout}~(c) on page~\pageref{fig:bnout}. A more informative scenario, scenario (d), has evidence entered into the various OOBN models to represent that the supplier who is applying for invoice financing is in a Golden Wait-a-Lot supply chain, has passed the credit check, is not a tier 1 supplier and the lower tier is funded by the same invoice finance company. In this situation the invoice financier is informed that the model predicts the most probable outcome to be to fund the supplier with a probability of 77.4\% and that the probability of the supply chain being stable is 76.8\% (Figure~\ref{fig:bnout2} on page~\pageref{fig:bnout2}). The final scenario excludes any information prior to the evidence entered, due to the structure of the model and only the evidence of two nodes are used to assess the stability of the supply chain that the supplier is part of: whether the decision was to fund the supplier and whether the invoices of another supplier in the supply chain that this supplier directly or indirectly supplies, is already being funded by the finance company. If the decision is not to fund the supplier, but the lower tier supplier is funded by the financier (scenario (e)), we see that the supply chain becomes highly unstable with 99.0\% probability.

It is important to note that the OOBN model structures and conditional probability tables were constructed to illustrate the interaction of various factors in invoice financing decision making, using information that can be retrieved using atomic crosschain function calls. As mentioned previously BN model development is an iterative process and the model structure and parameters refined over time. These probabilities were constructed in the context of the use case and we do not claim that they are accurate, but they can readily be updated and the model structure refined to suit specific requirements.

\subsection{Private blockchains and crosschain function calls}
\label{sec:xcresults}
\begin{figure*}
 \begin{center}
  \includegraphics[width=\linewidth,height=\textheight,keepaspectratio]{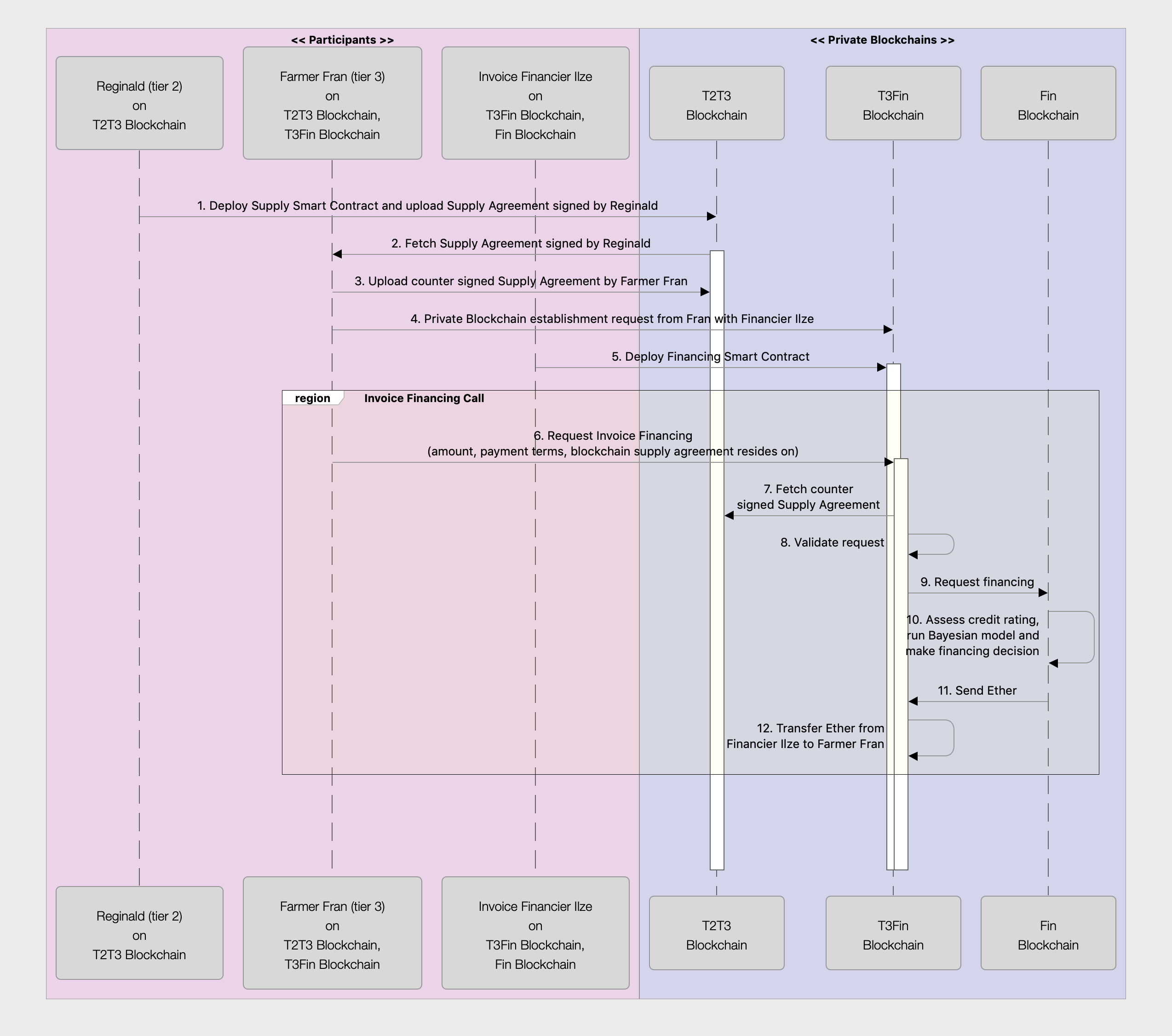}
  \caption{Invoice Financing Sequence Diagram}
  \label{fig:sequence}
 \end{center}
\end{figure*}

The sequence diagram in Figure \ref{fig:sequence} on page \pageref{fig:sequence} shows the private blockchain deployment, smart contract deployment, and atomic crosschain transaction flow. The participants of the system are listed on the left of the diagram, and the private blockchains to the right of the diagram. \\

Walking through the sequence diagram:
 
\begin{enumerate}

\item Regional Producer Reginald deploys a Supply Smart Contract to the T2T3 blockchain. 
He then uploads the Supply Agreement between Reginald and Farmer Fran, signed only by him, to the Smart Contract. \\

\item Farmer Fran downloads the Supply Agreement.\\

\item Farmer Fran counter-signs the Supply Agreement and uploads the counter-signed Supply Agreement to the Supply Smart Contract.\\

\item Farmer Fran establishes a private blockchain between herself and Invoice Financier Ilze. This private blockchain is called T3Fin blockchain.\\

\item Invoice Financier Ilze automatically deploys a standard Finance Smart Contract to the T3Fin blockchain immediately after the blockchain is deployed.\\

\item Farmer Fran submits an Atomic Crosschain Transaction to request invoice financing. In this request she specifies the required information: the amount of funding she is requesting, invoice payment terms, supply agreement terms, total amounts payable from unpaid invoices, and the blockchain on which the supply agreement resides.\\

\item The Atomic Crosschain Transaction call does a Crosschain Read, requesting the counter-signed Supply Agreement from the T2T3 blockchain. \\ 

\item The Atomic Crosschain Transaction performs some processing on the T3Fin blockchain to validate the invoice financing request from Farmer Fran. For example, it checks to see if the amount of funding requested is in agreement with the counter signed Supply Agreement.\\

\item The Atomic Crosschain Transaction passes the financing request to Invoice Financier Ilze's private blockchain, Fin blockchain, for processing.\\

\item The Atomic Crosschain Transaction then operates on the Fin blockchain. The credit rating of Farmer Fran is checked using the access that Invoice Financier Ilze has to the National Credit Reporting Bodies. Financier Ilze checks her customer list to see if Reginald who Fran has the supply agreement with is already being funded by her company.  The Bayesian network model is executed using whatever information is available to the financier, and the result of these analyses forms the basis for a funding decision.\\

\item Assuming the funding decision is to approve the funding for Farmer Fran, the Atomic Crosschain Transaction  results in Ether (money) being transferred from the Fin blockchain to the T3Fin blockchain.\\

\item The Atomic Crosschain Transaction then completes with the Ether (money) being transferred from Invoice Financier Ilze to Farmer Fran.\\

\end{enumerate}

\clearpage

\section{Discussion}
\label{sec:supplyprob}
The problem statement for the example use case, albeit fictitious, demonstrates that an upstream supplier (e.g. tier 2, 3 or 4) could encounter an unanticipated cashflow problem, despite conducting due diligence to ensure that they were in a good position to meet demand, prior to signing an initial supply agreement. The most astute and cost-effective finance option to address the cashflow problem is collatoralised finance, such as invoice financing. However, as discussed previously, it is not a simple task to obtain the required information to help secure funding. In fact, they typically face substantial hurdles with manufacturers and retailers not prepared to disclose sensitive business information, but wanting to keep this information private.

In some instances, for example if a financier is able to link the applicant and one of their customers as belonging to the same supply chain, then this information is important for invoice financiers to make well considered and informed funding decisions under uncertainty, by running a decision making tool such as a Bayesian network model which can generate the probability distribution across the states of the node that they are interested in, e.g. the perception of risk, or the decision to fund a supplier.

Tier 1 suppliers in a supply chain of a well known retailer, typically have no problem to have their invoices funded earlier by an invoicing financier, since they are considered as low risk companies and their approved invoices are read into the financier's systems, ready to be financed on the payment dates. Therefore if the tier 1 supplier offers the retailer a discount of say 5\% to fund their invoices ahead of the payment date, for reasons such as short term cash flow, or to pursue new business opportunities or to gain greater market share, this is a fairly straight forward transaction, which is most likely pre-approved. However, tier 2, 3, and 4 suppliers usually do not enjoy such ready access to funds. For example, Farmer Fran in the example use case would be a tier 3 supplier. In these instances the invoice financing company would not be interested to fund the supplier's invoices, unless further information is provided, which can be made accessible by crosschain function calls. In the Methods section on \pageref{sec:methods} we outline a potential solution combining novel technologies such as atomic crosschain functionality \cite{Robinson2019b} with Bayesian network models \cite{Pearl2014}. 

\section{Conclusion}
\label{sec:conclusion}
The proposed solution in this paper addresses the challenge of financing suppliers further upstream in the supply chain, who are often smaller suppliers, unable to demonstrate credit worthiness, possibly partly due to a lack of good accounts and documentation, and therefore unable to present the required information to secure funding. The crux of the matter is that if one or more of the participants in a supply chain are unable to get their invoices funded when necessary, then this is very likely to have repercussions for the rest of the participants in the supply chain, and potentially dire consequences for the financiers of the supply chain. 

Utilising the information obtained with atomic crosschain calls between private blockchains \cite{Robinson2019b} we have the ability to prove relationships in the supply chain without having to publicly disclose other more sensitive information  such as quantities and prices, and combined with BN modelling, an AI approach, we are able to gain access to information required by an invoice financier to make informed funding decisions under uncertainty. Moreover, the BN model structure and probabilities can be updated as more data become available and knowledge gaps are identified and addressed \cite{Johnson2012}. 

%
\section*{Acknowledgment}
This research was conducted while the authors were employed full-time at ConsenSys. We thank Horacio Mijail Anton Quiles and John Brainard for their insightful discussions, constructive critiques, and careful review, which have greatly benefitted the paper.
%
\bibliographystyle{IEEEtran}
\bibliography{IEEEabrv,sidechains2019-05-17}
%


\end{document}